\documentclass[%
 aip,
 groupedaddress, 
 amsmath,amssymb,
 reprint,%
]{revtex4-1}

\usepackage{graphicx}
\usepackage{dcolumn}
\usepackage{bm}

\usepackage[utf8]{inputenc}
\usepackage[T1]{fontenc}
\usepackage{mathptmx}

\begin{document}

\preprint{AIP/123-QED}

\title{Complete Stokes vector analysis with a compact, portable rotating waveplate polarimeter}

\author{T. A. Wilkinson}
    \email{taw0035@mix.wvu.edu}
\author{C. E. Maurer}
\author{C. J. Flood}
\author{G. Lander}
\author{S. Chafin}
\author{E. B. Flagg}
    \email{Edward.Flagg@mail.wvu.edu}
\affiliation{Department of Physics and Astronomy, West Virginia University, Morgantown, West Virginia 26506, USA}

\date{\today}

\begin{abstract}
Accurate calibration of polarization dependent optical elements is often necessary in optical experiments. A versatile polarimeter device to measure the polarization state of light is a valuable tool in these experiments. Here we report a rotating waveplate-based polarimeter capable of complete Stokes vector analysis of collimated light. Calibration of the device allows accurate measurements over a range of wavelengths, with a bandwidth of >30 nm in this implementation. A photo-interrupter trigger system supplies the phase information necessary for full determination of the Stokes vector. An Arduino microcontroller performs rapid analysis and displays the results on a liquid crystal display. The polarimeter is compact and can be placed anywhere on an optical table on a single standard post. The components to construct the device are only a fraction of the cost of commercially available devices while the accuracy and precision of the measurements are of the same order of magnitude.
\end{abstract}

\maketitle


The polarization state of light, described by the Stokes vector $\textbf{S} = [S_0, S_1, S_2, S_3]$ \cite{stokes_mathematical_1880}, is of critical importance in a broad range of optics-based experiments \cite{foreman_information_2010}. Coupling light to a polarization maintaining fiber requires that the polarization of the input beam be matched to the polarization axis of the fiber \cite{noda_polarization-maintaining_1986}. Selectively addressing optically active transitions in quantum emitters like atoms or quantum dots requires the polarization state of the incident laser match the dipole moment of the desired transition \cite{wilkinson_spin-selective_2019}. Calibration of liquid crystal variable retarders (LCVRs), a common polarization manipulation component, necessitates precise knowledge of the polarization of the laser to be used as a reference \cite{bueno_polarimetry_2000}. These are among many other experimental situations that need highly accurate knowledge of the polarization state of light. Therefore, the complete measurement and characterization of the polarization of light is of widespread interest and importance.

Here we describe a self-contained, portable, inexpensive polarimeter capable of complete Stokes vector characterization of any input polarization. The polarimeter is based on a rotating quarter-wave ($\lambda$/4) plate (QWP), a polarizer (called the analyzer), and a photodiode in series \cite{berry_measurement_1977}. In general, linearly polarized light can be distinguished from circularly polarized light by analyzing the fluctuating light intensity signal incident on the photodiode. This approach relies on measuring the relative amplitudes of the 2$\omega$ and 4$\omega$ components of the signal, where $\omega$ is the angular frequency of the rotating QWP in front of the analyzer \cite{bobach_note:_2017,hauge_survey_1976}. However, to fully measure the Stokes vector, the relative angle between the QWP fast axis (FA) and the analyzer transmission axis (TA) must be known for each period of rotation \cite{romerein_calibration_2011}. We use a photo-interrupter and a timing disk to measure the phase of the QWP angle relative to the analyzer for each rotation period. We combine our optical setup with an Arduino microcontroller for analysis. Our implementation is thus able to measure any arbitrary input polarization, while resting on only a single standard optical post mount. The polarimeter presented here provides advantages in ease of use, functionality, portability, and cost over previous realizations \cite{bobach_note:_2017,romerein_calibration_2011,kihara_measurement_2011}.

\begin{figure*}[t]
    \centering
    \includegraphics{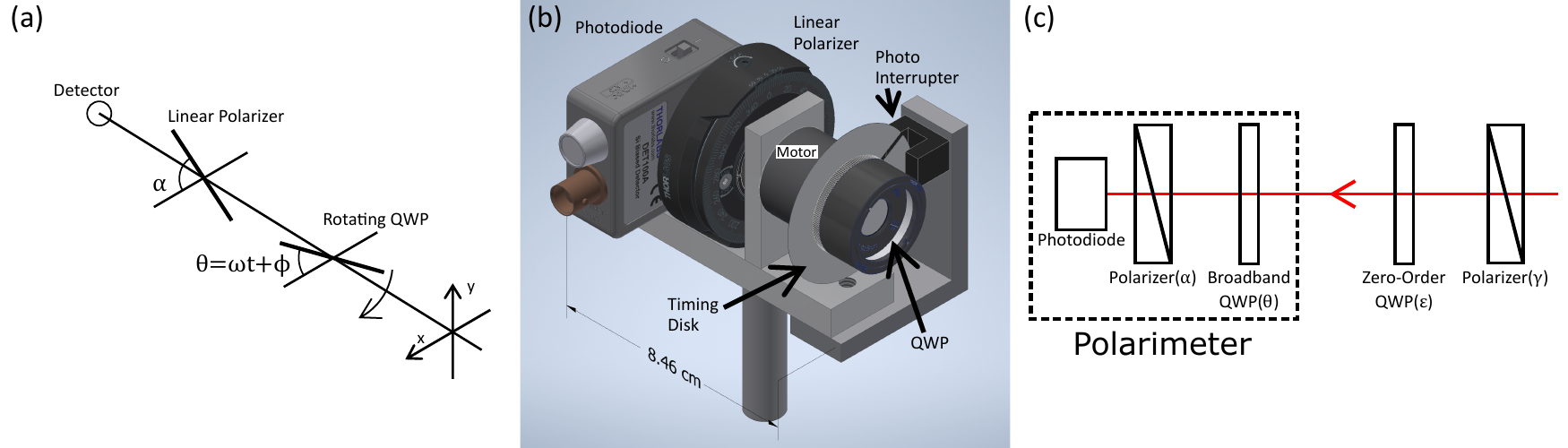}
    \caption{Measurement part of the polarimeter shown (a) schematically, and (b) as a 3-D image of the physical device. The QWP is mounted on a hollow axle motor with azimuth $\theta = \omega t + \phi$, where $\omega$ is the angular frequency of rotation and $\phi$ the phase of the rotation, and $\alpha$ is the azimuth of the linear analyzer. (c) The optical setup for calibration and characterization of the device. The upstream polarizer has azimuth $\gamma$, and the upstream QWP has azimuth $\epsilon$.}
    \label{fig:schematic}
\end{figure*}

The polarization state of a beam of light can be formally described by the Stokes vector $\textbf{S} = [S_0, S_1, S_2, S_3]$, where $S_0=I_X+I_Y$ is the Stokes parameter describing the total intensity, $S_1=I_X-I_Y$ the linear component, $S_2=I_D-I_A$ the diagonal component, and $S_3=I_R-I_L$ the circular component, where $I_\beta$ with $\beta=\{X,Y,D,A,R,L\}$ denotes 
the measured intensity of a given polarization
\cite{stokes_mathematical_1880}. The manipulation of the polarization by a series of optical elements can be described by the appropriate (4x4) Mueller matrices \cite{hecht_optics_2017,anderson_measurement_1992,compain_general_1999}. Figure~\ref{fig:schematic}(a) shows the elements of our system, namely a rotating waveplate of retardance $\delta\approx\pi/2$ at angle $\theta$, followed by a stationary linear polarizer at angle $\alpha$ before the detector. Operating the Mueller matrices of the optical components on an arbitrary input Stokes vector, one arrives at a general expression for the intensity signal detected by the photodiode as \cite{berry_measurement_1977,bobach_note:_2017,schaefer_measuring_2007}
\begin{equation}
\label{eqn:intensity}
    I(\theta) = \frac{1}{2}\left[A + B\sin(2\theta) + C\cos(4\theta) + D\sin(4\theta)\right],
\end{equation}
\noindent where $\theta = \omega t + \phi$ is the orientation angle of the waveplate, with $\omega$ the angular frequency of rotation and $\phi$ describing the phase. Allowing for an imperfect retardance of the waveplate ($\delta \ne \pi/2$) the coefficients of the frequency components of Eq.~\ref{eqn:intensity} give the input Stokes parameters by the relations \cite{flueraru_error_2008, arnoldt_rotating_2011, kihara_measurement_2011}
\begin{subequations}
\label{eqn:components}
\begin{align}
    S_0 &= A-C/\tan^2(\delta/2), \\
    S_1 &= C/\sin^2(\delta/2), \\
    S_2 &= D/\sin^2(\delta/2), \\
    S_3 &= B/\sin(\delta).
\end{align}
\end{subequations}

The relationships of the coefficients in Eq.~\ref{eqn:intensity} to different frequency sinusoids means the Fourier transform provides a natural tool for determination of the Stokes vector. The linear components of the Stokes vector ($S_1$, $S_2$) depend only on the $4\omega$ part of the signal, while the circular component ($S_3$) depends only on the $2\omega$ part. Using this fact, one can easily determine the degree of linear or circular polarization of the light without knowing the phase of the signal, as in the implementation by Bobach et al. \cite{bobach_note:_2017}. However, to measure the complete Stokes vector of the light, the components $C$ and $D$ must be separated, and thus the phase of the signal ($\phi$) must be known.

The angle $\phi$ determines the position of the FA of the QWP at the beginning of each period of rotation. In principal, the signal may be analyzed and the Stokes vector determined for any choice of analyzer angle $\alpha$ and phase angle $\phi$. However, situations where the analyzer does not lie in the desired measurement basis ($\alpha \ne 0, \pi/2$), or the QWP FA is not parallel or perpendicular to the analyzer at the start of a new period ($\lvert \alpha - \phi \rvert \ne 0, \pi/2$), lead to unnecessary complications in Eq.~(\ref{eqn:intensity}) and the subsequent Fourier analysis. Thus, we will limit ourselves to cases where $\alpha = 0, \pi/2$ and $\lvert \alpha - \phi \rvert = 0, \pi/2$. The remaining choices of analyzer angle $\alpha$ and QWP trigger angle $\phi$, of which there are four, affect the signs in Eq.~(\ref{eqn:intensity}). For $\alpha = \phi = 0$, Eq.~\ref{eqn:intensity} is the result, while for the choice $\alpha = 0$, $\phi = \pi/2$ the second term becomes negative and the equation matches those found in \cite{flueraru_error_2008} and \cite{arnoldt_rotating_2011}. In our current implementation we chose the case of Eq.~(\ref{eqn:intensity}) where $\alpha = \phi = 0$.

Figure~\ref{fig:schematic}(b) shows a 3-D image of our experimental measurement apparatus. As in the previous implementation by Bobach et al. \cite{bobach_note:_2017} we mount the QWP on a model airplane motor with a hollow axle, which allows light to pass through the axle to the analyzer and detector. Attached to the rotating part of the motor we place a thin, circular metal timing disk with a slit cut out near the edge and a hole in the center. A photo-interrupter is positioned around the timing disk and produces a trigger signal each time the small slit passes through. The combination of the timing disk and the photo-interrupter serves as a trigger for the start of each period of rotation of the QWP.

\begin{figure*}[t]
    \centering
    \includegraphics{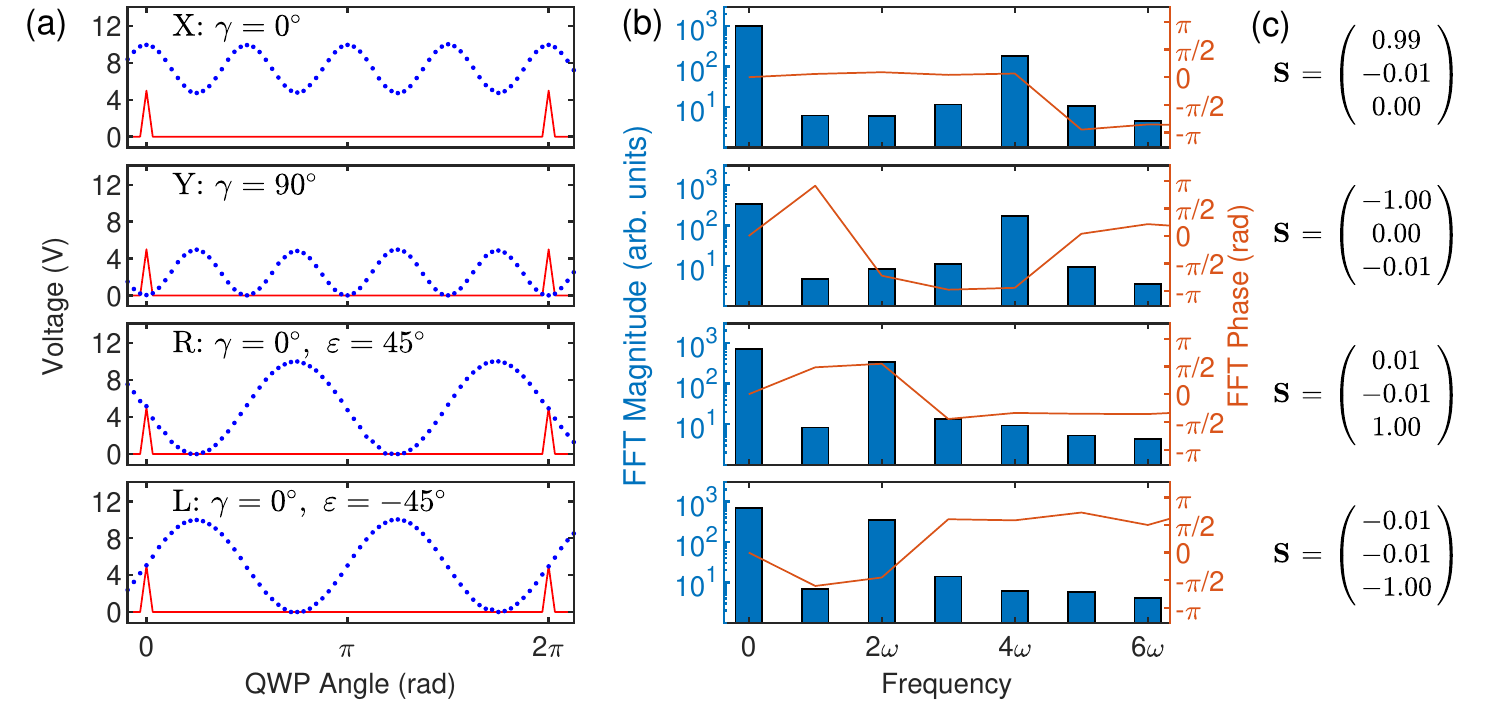}
    \caption{(a) Measured photodiode voltage (points) and trigger signal (line), (b) FFT magnitude and phase, and (c) calculated Stokes vectors for four input polarizations: X, Y, R, and L. The angular orientations of the upstream optics are noted in (a), with $\gamma$ the angle of the polarizer and $\epsilon$ the angle of the zero-order QWP. These data were taken at a wavelength of 912.9340 nm, the measured $\lambda$/4 value of the upstream QWP.}
    \label{fig:FFT}
\end{figure*}

Physically achieving the desired angle of the QWP FA when the trigger fires ($\phi = 0$) requires a small amount of calibration of the device. For a known X-polarized input ($\textbf{S} = [1, 1, 0, 0]$) the longitude angle of the Stokes vector on the Poincaré sphere, $2\psi=\arctan(S_2/S_1)$, should be zero. We use this longitude as feedback for the calibration. In principle this can be achieved entirely by hand with a careful alignment of the QWP in its mount. In practice this proves difficult and time consuming due to the lack of fine control of our custom-made QWP mount designed for use with the hollow axle motor (design in supplemental). So we add in software the ability to offset the measured signal by $360^\circ/64 = 5.625^\circ$, where 64 is the number of data points sampled from the photodiode per period (see Fig.~\ref{fig:FFT}(a)). Using these two methods in tandem to set $2\psi = (0 \pm 1)^\circ$ for an X-polarized input calibrates the device quickly and accurately. Note that the software offset allows the physical phase angle $\phi$ to differ from zero. However, this difference leads only to a phase shift of the signal that is corrected by the software offset. This method essentially allows misalignment of $\phi$ in increments of $5.625^\circ$, rather than specifically requiring $\phi = 0$.

The voltage signals from the photodiode and the photo-interrupter are sent to a small circuit board containing an Arduino Pro-Micro controller for signal processing. The user can control the duration of the software offset using a potentiometer connected to an analog input of the Arduino. Adjustment of the offset combined with physical realignments of the QWP in its mount form the basis of our calibration procedure. The Arduino performs the fast Fourier transform (FFT), calculates the Stokes vector components, and displays the results on a small liquid crystal display (LCD). The circuit diagram, a detailed description of the calibration procedure, and the Arduino code are included in the supplemental.

The measurement head of the polarimeter shown in Fig.~\ref{fig:schematic}(b) measures about $7\times9\times7$ cm, and is conveniently mounted on a single standard optical post. The analysis part of the device containing the Arduino controller and circuit are housed in a small plastic box measuring $16\times16\times4$ cm. The two parts are connected by a single D-sub cable $\sim$3 m in length. Note that due to the induced electro-magnetic field (EMF) from the motor signal it is necessary to isolate the photo-interrupter wires from the rest of the wires in the cable. The setup is compact and can be easily placed anywhere on an optical table using a standard post holder. Additionally, the components in total cost less than \$1500, much of which is the cost of the QWP. Thus, our polarimeter is far less expensive than commercially available devices (e.g., Thorlabs PAX1000 series: \$5,718.25). A complete bill of materials can be found in the supplemental.

Accurate calibration and characterization of the device necessitates the production of known polarizations. We use a simplified version of the RLB method \cite{romerein_calibration_2011,brooks_polarization_1978}, which implements iterative rotations about the vertical axis for calibration of the azimuth of an optic to the vertical or horizontal axis. A $180^\circ$ rotation of an optic around the vertical axis has no effect on the transmitted light intensity if an optical axis is aligned either horizontally or vertical. Therefore iterative vertical rotations combined with rotation of the optic in its mount lead to calibration of either the transmission axes of a polarizer or the fast axis of a wave plate. This method has the benefit of not needing an already calibrated polarization optic as reference. An explicit description can be found in the supplemental. 

In principle circularly polarized light can be produced with a polarizer and a QWP. In practice, however, the retardance $\delta$ of a QWP depends largely on the wavelength $\lambda$ of the light. Thus, we measure the retardance as a function of wavelength using the method of Wang et al.~\cite{wang_novel_2004,wang_method_2006} for two QWPs: one zero-order wave plate with nominal retardance $\lambda/4$ at 915 nm, and one broadband wave plate with a specified operating range of 690-1200 nm. The zero-order wave plate has a linear dependence in the area around $\lambda = 915$ nm, and is measured to be truly quarter-wave ($\delta=\pi/2$) at $\lambda$ = 912.9340 nm. This enables production of circularly polarized light at this wavelength. The broadband wave plate is seen to have little dependence on wavelength over a 30 nm range around 915 nm, and we measure a retardance $\delta = (0.5167 \pm 0.0008) \pi$ over that range. We use the broadband QWP in the polarimeter and use Eq.~\ref{eqn:components} to compensate for the imperfect retardance. This enables accurate measurements over a range of wavelengths fully covering our specific application. This range could be expanded by measuring the retardance of the broadband QWP over a larger range.

Figure~\ref{fig:schematic}(c) shows the optical setup used to evaluate the performance of the polarimeter. A calibrated linear polarizer at angle $\gamma$, and the calibrated zero-order QWP at angle $\epsilon$ are placed upstream of the polarimeter to produce varying input polarizations. For all measurements described below, the wavelength of the input laser light was 912.9340 nm, corresponding to the measured $\lambda/4$ value of the zero-order QWP. Figure~\ref{fig:FFT}(a) shows the voltage signal output by the photodiode and the trigger signal generated by the photo-interrupter for four input polarizations. Each case is labeled with the intended input polarization and the angles $\gamma$ and $\epsilon$ used to produce it. For the X and Y inputs the QWP was not in the optical path. Figure~\ref{fig:FFT}(b) shows the calculated FFT magnitude and phase for the photodiode signal; note that the magnitude is plotted on a logarithmic scale. The linear cases (X, Y) show large $4\omega$ components and the circular cases (R, L) show large $2\omega$ components, both as expected. The phase of the $4\omega$ component for X is near 0, while for Y it is near $-\pi$. The same $\pi$ phase difference can be seen for the $2\omega$ components in the circular (R, L) cases. This shows explicitly how these polarizations are indistinguishable from each other without access to the phase information that our photo-interrupter trigger system provides. Figure~\ref{fig:FFT}(c) shows the calculated Stokes vector for each polarization. Note that we omit $S_0$ in the vector here as it is simply used to normalize the other components. The values are rounded in congruence with their appearance on our LCD. In each case, every component of the measured Stokes vector is within 1\% of the expected value.

\begin{figure}[t]
    \centering
    \includegraphics{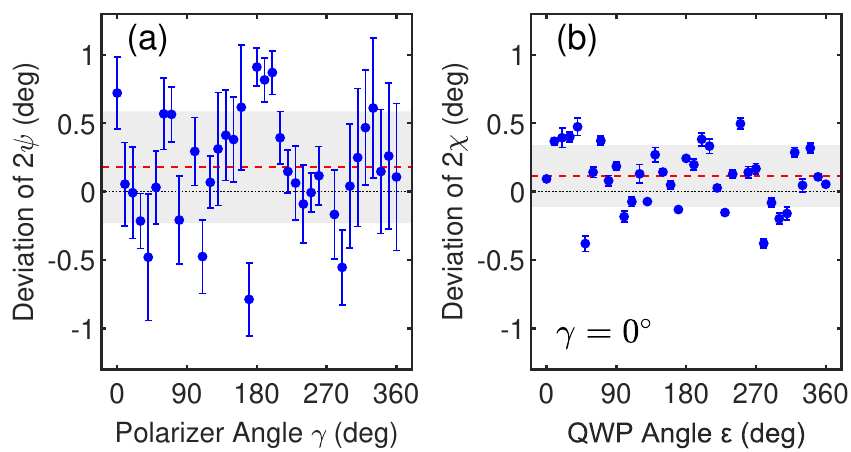}
    \caption{Deviation from expected value for (a) the longitude angle $2\psi$ and (b) the elevation angle $2\chi$, when the upstream optics are rotated. In (a) the polarizer angle $\gamma$ is rotated; in (b) the zero-order QWP angle $\epsilon$ is rotated. The dashed red line is the mean of the data, and the shaded region is $\pm1$ standard deviation from the mean.}
    \label{fig:deviations}
\end{figure}

The Stokes vectors measured and reported in Fig.~\ref{fig:FFT}(c) show that the device is working properly for those input polarizations. To fully evaluate the device for all input polarizations we rotate the angles of our upstream optics (Fig.~\ref{fig:schematic}(c)) and compare the measured angles of the Stokes vector on the Poincaré sphere with those expected. To evaluate linear polarizations we omit the zero-order QWP and rotate the polarizer angle $\gamma$ while measuring the deviation from the expected value of the longitude angle $2\psi$ defined previously. These data are shown in Fig.~\ref{fig:deviations}(a) with the mean of the data shown as a dashed red line and $\pm1$ standard deviation from the mean as a shaded region. The standard deviation is $0.41^\circ$ and the mean is $0.18^\circ$, which is within half a standard deviation from zero. To evaluate linear, circular, and elliptical polarizations all in tandem we leave the polarizer at $\gamma = 0^\circ$ while rotating the zero-order QWP angle $\epsilon$ and measuring the deviation from the expected value of the elevation angle $2\chi = \arctan\left(S_3 / \sqrt{S_1^2 + S_2^2}\right)$. These data are shown in Fig.~\ref{fig:deviations}(b), with the same representation of the mean and standard deviation. Here we obtain a standard deviation of $0.23^\circ$ and a mean of $0.11^\circ$, again within half a standard deviation from zero. We attribute the difference in the observed uncertainties of the individual measurements of the two angles to the differing dependencies of the angles on the frequency components of the intensity signal defined in Eqs.~\ref{eqn:intensity} and~\ref{eqn:components}. Specifically, the longitude angle $2\psi$ depends on the complex phase of the 4$\omega$ component, while the elevation angle $2\chi$ depends on the relative magnitudes of the 2$\omega$ and 4$\omega$ components, which is more stable. The mean and standard deviation values measured here show quantitatively that the accuracy of the polarimeter is near that of commercially available devices. For example, the PAX1000 series of polarimeter from Thorlabs has a specified accuracy of $\pm0.25^\circ$.

We report a self-contained polarimeter capable of fully characterizing the Stokes vector to within one degree on the Poincaré sphere. Our device provides accuracy comparable to the leading commercial devices for a fraction of the cost without sacrificing size and ease of use as seen in previous non-commercial realizations \cite{bobach_note:_2017,romerein_calibration_2011,flueraru_error_2008}. By using a photo-interrupter as a trigger and an Arduino micro controller to perform the calculations and display the results on an LCD, we provide a compact, user-friendly, and cost effective way to quickly and accurately measure the polarization of collimated light.

\section*{Supplementary Materials}
See supplementary materials for supporting information about construction of the device.

\begin{acknowledgments}
This research was supported by the U.S. Department of Energy, Office of Basic Energy Sciences, Division of Materials Sciences and Engineering under Award No. DE-SC0016848. T. A. Wilkinson acknowledges support from the Research Corporation for Science Advancement.
\end{acknowledgments}

\section*{AIP Publishing Data Sharing Policy}
The data that supports the findings of this study are available within the article.


\bibliography{PolarimeterReferences}

\end{document}